\documentclass[fleqn,10pt]{wlscirep}
\usepackage[pdftex]{dropping}

\title{Cell-Type-Specific Differences in KDEL Receptor Clustering in 
Mammalian Cells}

\author[1]{Achim Bauer}
\author[2]{Ludger Santen}
\author[1]{Manfred J. Schmitt}
\author[2,*]{M.\ Reza Shaebani}
\author[1,*]{Bj\"orn Becker}
\affil[1]{Molecular and Cell Biology, Department of Biosciences 
and Center of Human and Molecular Biology (ZHMB), Saarland 
University, 66123 Saarbr\"ucken, Germany}
\affil[2]{Department of Theoretical Physics and Center for Biophysics, 
Saarland University, 66123 Saarbr\"ucken, Germany}
\affil[*]{equal contribution.}
\affil[ ]{Correspondence should be addressed to B.\,B.\ 
(bjoern\_becker2@gmx.de) or M.\,R.\,S.\ (shaebani@lusi.uni-sb.de).}

\begin{abstract}
In eukaryotic cells, KDEL receptors (KDELRs) facilitate the retrieval of 
endoplasmic reticulum (ER) luminal proteins from the Golgi compartment 
back to the ER. Apart from the well-documented retention function, recent 
findings reveal that the cellular KDELRs have more complex roles, e.g.\ 
in cell signalling, protein secretion, cell adhesion and tumorigenesis. 
Furthermore, several studies suggest that a sub-population of KDELRs is 
located at the cell surface, where they could form and internalize KDELR/cargo 
clusters after K/HDEL-ligand binding. However, so far it has been unclear 
whether there are cell-type- or species-specific differences in KDELR 
clustering. By comparing ligand-induced KDELR clustering in different 
mouse and human cell lines via live cell imaging, we show that macrophage 
cell lines from both species do not develop any clusters. Using RT-qPCR 
experiments and numerical analysis, we address the role of KDELR expression 
as well as endocytosis and exocytosis rates on the receptor clustering 
at the plasma membrane and discuss how the efficiency of directed transport 
to preferred docking sites on the membrane influences the exponent of 
the power-law distribution of the cluster size.
\end{abstract}

\begin{document}

\flushbottom

\maketitle

\thispagestyle{empty}


\section*{Introduction}

Recent discoveries in the KDEL receptor (KDELR) research field 
have strongly changed the common understanding of the role of these fascinating 
transmembrane proteins. It is obvious now that the three KDELR homologues have 
more diverse and fundamental isoform-specific roles in eukaryotic systems than 
previously assumed \cite{Capitani09,Raykhel07}; KDELRs do not merely 
maintain the composition of the endoplasmic reticulum (ER) by returning 
ER-resident proteins from the Golgi into the ER via a pH-dependent retrieval 
mechanism \cite{Brauer19,Semenza90,Wilson93,Lewis90}. Under stress condition, 
KDELR2 and KDELR3 expression is upregulated on the transcriptional level via 
the Xbp1/Ire1 pathway in mammalian cells to counteract the loss of ER-resident 
proteins \cite{Trychta18}. Previous studies also indicated that KDELRs regulate 
Golgi homeostasis as well as protein secretion by interacting with a subset 
of different G-proteins at the Golgi membrane \cite{Giannotta12,Pulvirenti08,
Tapia19}. After KDELR/ligand interaction in the Golgi lumen, the active 
G$\alpha$ subunits activate their specific target protein kinases (e.g.\ 
Src-kinases or PKA), which subsequently modulate gene transcription followed 
by regulation of the anterograde or retrograde trafficking \cite{Giannotta12,
Pulvirenti08,Sallese09}. The regulation of protein secretion is mediated by a 
cellular mechanism, called ``traffic-induced degradation response for secretion" 
(TIDeRS), which activates KDELR1-dependent PKA signalling and results in a complex 
interplay between the cytoskeleton, autophagy and secretion machinery including 
lysosome relocation as well as autophagy-dependent lipid-droplet turnover 
\cite{Tapia19}. KDELR malfunctions are associated with changes in extracellular 
matrix degradation and cellular adhesion \cite{Blum19,Ruggiero15,Ruggiero17}. 
Recent studies have revealed that upregulated KDELR2 expression level in 
glioblastoma tissues promotes the tumorigenesis and shortens the lifetime, 
making the receptor an interesting therapeutic target in glioblastoma patients 
\cite{Liao19}.  

It is known that a sub-population of KDELRs in mammalian and yeast cells are 
located at the cell surface \cite{Becker16,Becker16b,Henderson13}, however, 
the possible reasons of this plasma membrane (PM) localization are not fully 
understood. It is suggested that the transport of the ER chaperone isoform 
PDIA6 to the cell surface depends on its KDEL-motif and is presumably mediated 
by KDELR1 \cite{Bartels19}. Also, PM-localized KDELRs in \emph{S.\ cerevisiae} 
serve as specific A/B toxin receptors which are hijacked by the yeast killer 
toxin K28 to ensure its cell entry \cite{Becker16}. Nevertheless, a more 
natural role of KDELRs at the yeast cell surface is the reinternalization 
of mistrafficked ER-resident proteins from the yeast PM to prevent their 
permanent loss as well as their new synthesis, thus, saving energy and cellular 
resources \cite{Becker16}. Based on recent studies using mesencephalic 
astrocyte-derived (MANF) or cerebral dopamine (CDNF) neurotrophic factors 
\cite{Henderson13,Maciel19}, it seems that the KDELRs at the cell surface 
are also involved in cell-cell communication by sensing ER stress level between 
tissue cells through binding secreted KDELR ligands of stressed neighbouring 
cells at the PM level. For CDNF and MANF, a neuro- and cardio-protective effect 
is postulated which is presumably mediated via initial KDEL receptor binding 
and upstream activation of signalling pathways (e.g.\ PI3K/AKT for CDNF), 
similar to KDELR-dependent signalling processes at the Golgi compartment 
\cite{Henderson13,Maciel19}. Cells have surely developed specific mechanisms 
and/or signalling pathways to respond to ligand binding and to regulate the 
KDELR expression level at the cell surface, however, the molecular machinery 
responsible for KDELR transport to the PM has not been well-characterized. 

So far, it is known that KDELRs form clusters in HeLa cells in the presence 
of an artificial model cargo containing a C-terminal ER retention motif (HDEL 
or KDEL) \cite{Becker16b}. It has also been demonstrated that cargo binding 
induces an increased microtubule-assisted KDELR transport to preferred arrival 
sites  at the PM \cite{Becker16b}. However, it is unclear whether cluster 
formation at the PM is a cell-type- or species-specific process. Here we study 
KDELR clustering in different mouse and human cell lines by live cell imaging. 
Our main observation is that mouse and human macrophage cell lines do not 
show any KDELR cluster formation at the cell surface after ligand treatment. 
Additionally, the clustering dynamics is qualitatively similar in cell types 
which develop receptor clusters at the PM, independent of species identity. 
By means of RT-qPCR experiments, we exclude the possibility that the low mRNA 
level of KDELRs is responsible for the missing receptor clustering phenotype 
seen in macrophages. We consider a stochastic KDELR endo/exocytosis model 
and perform Monte Carlo simulations to better understand how the differences 
in cluster formation in various cell types may originate from the differences 
in their endocytosis and/or exocytosis rates. 

\section*{Methods}

\subsection*{Cultivation of human and mouse cell lines} 
HeLa (ATCC number CCL-2), HEK-293T (Invitrogen), SH-SY5Y (Sigma), RAW-Blue 
(Invitrogen), L929 and MEF cells were cultivated in DMEM medium (Gibco) 
supplemented with $1\%$ penicillin/streptomycin (PAA) and $10\%$ fetal 
bovine serum (Biochrom) in a humidified environment at $37^{\circ}$C 
and $5\%$ $\text{CO}_{2}$. IC-21 and THP1 (ATCC number TIB-202) were 
cultivated in RPMI-1640 medium (Gibco) supplemented and cultivated as 
listed above. To differentiate THP1 cells to macrophages, cells were 
pre-treated with phorbol 12-myristate 13-acetate (PMA, 30$\,\mu$g/ml) 
for 72 h and subsequently used for live cell imaging. 

\subsection*{Production/purification of KDELR model cargo} 
Expression of enhanced GFP-tagged RTA variants $\text{eGFP-RTA}^\text{E177D}$ 
and $\text{eGFP-RTA}^\text{E177D-HDEL}$ in {\it E.\ coli} and the subsequent 
affinity purification procedure was performed as previously described 
in \cite{Becker16b}. Substitution of aspartate for glutamate at position 
177 in the model cargo leads to a 50 fold reduction of RTA cytotoxicity 
\cite{Li10} and was done by conventional PCR with primers listed in {\it 
Suppl.\,Table\,S1}. 

\subsection*{Live cell imaging} 
In imaging experiments, $1.5{\times}10^5$ cells of different cell lines 
were seeded out in $60\mu$-ibiTreat-dishes (Ibidi) and pre-cultivated for 
24 h. Next, the cells were washed two times with PBS (pH 7.4) and cultivated 
in DMEM (w/o phenol red, $10\%$ FCS) and subsequently analyzed by confocal 
laser scanning microscopy. To investigate cargo-induced clustering at the 
PM, cells were treated with 160$\,\mu$g/ml of $\text{eGFP-RTA}^\text{E177D-HDEL}$ 
and monitored for 3 h at $37^{\circ}$C and $5\%$ $\text{CO}_{2}$. Thereby, 
the RTA variant lacking a KDELR binding site ($\text{eGFP-RTA}^\text{E177D}$, 
160$\,\mu$g/ml) served as negative control and was monitored for 1.5 h. The 
time resolution in each experiment is expressed as frames per hour (frames/h). 

\subsection*{Confocal microscopy} 
Live cell imaging of $\text{eGFP-RTA}^\text{E177D}$ or $\text{eGFP-RTA}^
\text{E177D-HDEL}$ was performed by confocal fluorescence microscopy using 
a Zeiss LSM 510 META (Nikon PlanApo 63x NA 1.4 oil immersion lens, 488 nm 
excitation, $1.1\%$ argon laser power, HFT 488 and NFT 490 beam splitter, 
BP 500-530 filter). The same laser power and pinhole size (73 $\mu$m) were 
used to collect all images in each experiment. 

\subsection*{Evaluation of cluster-size distribution} 
The frames extracted from the experimental videos were converted to 
gray scale images. Next, an anisotropic Gaussian filter was used to 
smoothen the intensity field by determining the background noise 
around each local intensity peak. A threshold ratio between each 
local peak and its background intensity was used to subtract the 
noise to obtain the receptor clusters. The chosen parametrization 
of the smoothening procedure does not qualitatively influence the 
image analysis results. By converting the pixel gray-scale intensity 
to a binary array and using the Hoshen-Kopelman algorithm the 
clusters were identified and by a logarithmic binning of the size 
range the cluster-size distribution was obtained. 

\subsection*{Gene expression analysis via real time qPCR } 
For RNA isolation, $1{\times}10^6$ cells were cultivated in the appropriate 
medium for 24 h in 6-well plates and total cellular RNA was isolated 
using the Direct-zol RNA MiniPrep Plus Kit (Zymo Research) following 
the manufacturer's instructions. Next, 500 ng of RNA was transcribed 
into cDNA using Maxima Reverse Transcriptase (200 U, Thermo Fisher 
Scientific), Oligo(dT)18 Primer (100 nM, Thermo Fisher Scientific) 
and dNTP Mix (each dNTP 0.4 mM, Thermo Fisher Scientific). Finally, 
a 10 $\mu$l qPCR mix was prepared including 2 ng cDNA, 0.2-1 mM of 
the corresponding primers (see {\it Suppl.\,Table\,S2}) and 2 $\mu$l 
of 5 $\times$ Hot-Start-Taq2 qPCR EvaGreen Mix (Axon). RT-qPCR was 
performed in 40 amplification repeats according to the Hot-Start-Taq2 
qPCR EvaGreen manual instructions with primer-optimized annealing 
temperature using CFX Connect Real-Time System (BioRad). Data 
analysis was carried out with CFX Manager 3.1 (BioRad). 

\subsection*{Primers, probes and statistical analysis} 
KDELR primer efficiency was analyzed in a DNA standard curve by a 
five-log dilution series of either HeLa or L929 cDNA. A no-template 
control or no-reverse transcriptase control was applied to detect 
genomic DNA contaminations. Biological replicates ($n{=}3$) as well 
as technical replicates ($n{\geq}2$) were used to determine KDELR 
gene expression levels. All Cq values were normalized to the mean 
of the Cq values of the reference gene glyceraldehyde 3-phosphate 
dehydrogenase (GAPDH) and are represented as a mean for $\text{E}
^\text{-dCq}$ (E=primer efficiency) or $\text{E}^\text{-ddCq}$ 
with error bars representing the upper and lower limits based on 
the standard deviation of delta Cq values. Statistical analysis 
was carried out in Graphpad Prism8. All pooled data were given 
as mean values $\pm$ SEM, and statistical significance was assessed 
by one-way ANOVA based on biological replicates and at sample 
sizes of $n{=}3$. 

\section*{Results}
\subsection*{KDELR clustering is cell-type specific and independent 
of species identity.} 
\begin{figure}
\centerline{\includegraphics[width=0.99\textwidth]{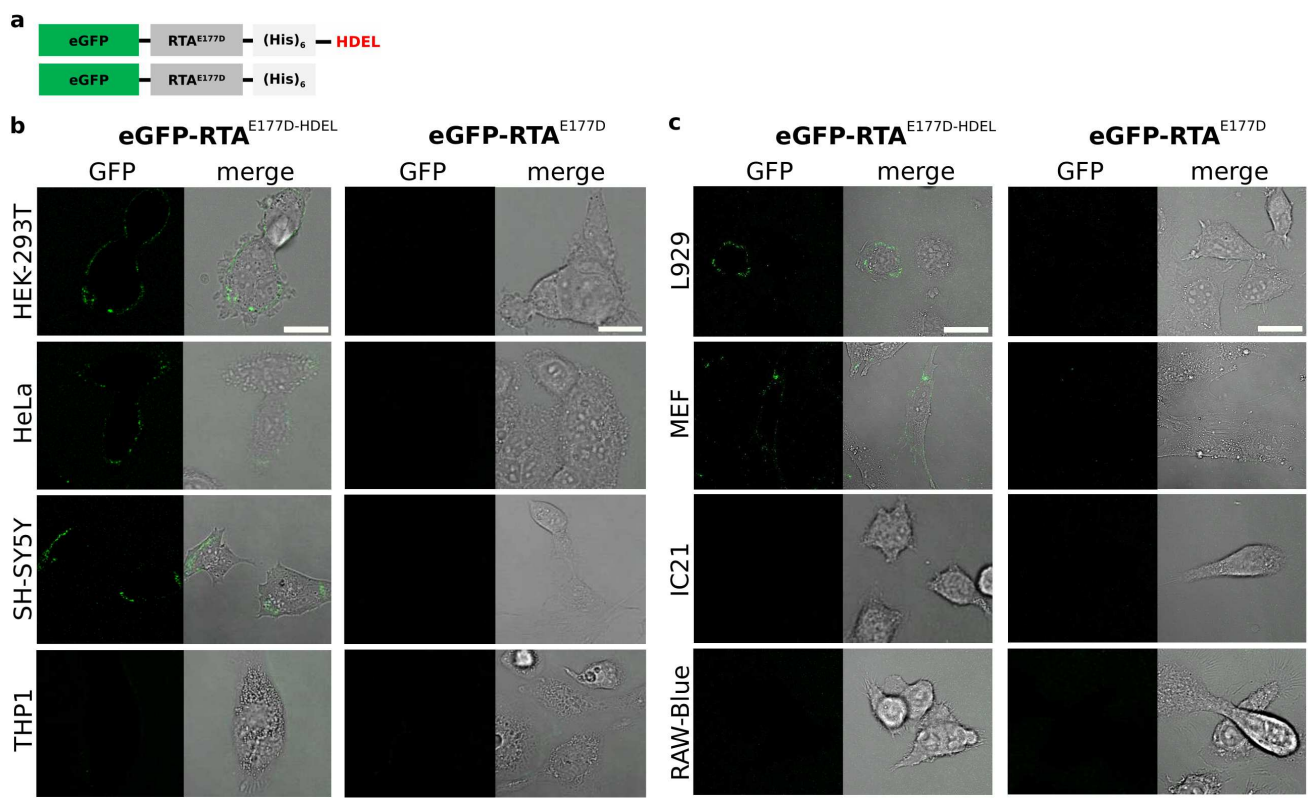}}
\caption{(a) (top) Schematic of the fluorescent model cargo 
$\text{eGFP-RTA}^\text{E177D-HDEL}$ consisting of the mammalian enhanced 
GFP (eGFP), a less toxic variant of RTA ($\text{RTA}^\text{E177D}$), 
and a C-terminal $\text{(His)}_6$-Tag for purification. (bottom) 
$\text{eGFP-RTA}^\text{E177D}$ lacking a KDELR binding motif, which 
is served as negative control. (b) Confocal laser scanning microscopy 
of human cell lines treated with $\text{eGFP-RTA}^\text{E177D-HDEL}$ or 
$\text{eGFP-RTA}^\text{E177D}$ (negative control). In the latter case, 
the images represent the steady-state regime of receptor cluster 
development ($t\geq150\,\text{min}$). Scale bars, 20 $\mu$m. (c) 
Similar to panel (b) but for mouse cell lines.}
\label{Fig1}
\end{figure}

Previous studies on HeLa cells have demonstrated a ligand-induced KDELR 
cluster formation at PM-localized KDELR hot spots \cite{Becker16b}. In 
the present work, we perform live cell imaging experiments in various 
cell lines from human (HeLa, HEK-293T, THP1 and SH-SY5Y) as well as 
mouse (L929, MEF, RAW-Blue and IC21) to determine cell-type- and/or 
species-specific differences in KDELR cluster formation after external 
ligand application. In order to minimize the cytotoxicity-induced side 
effects of the cargo, we replace the wild-type A-subunit of ricin (RTA) 
of the GFP-tagged model cargo $\text{eGFP-RTA}$ and $\text{eGFP-RTA}^\text{HDEL}$ 
(used in the previous study \cite{Becker16b}) with a 50-70-fold less toxic 
$\text{RTA}^\text{E177D}$ variant \cite{Li10} (see Fig.\,\ref{Fig1}a). 

\begin{figure*}
\centerline{\includegraphics[width=0.99\textwidth]{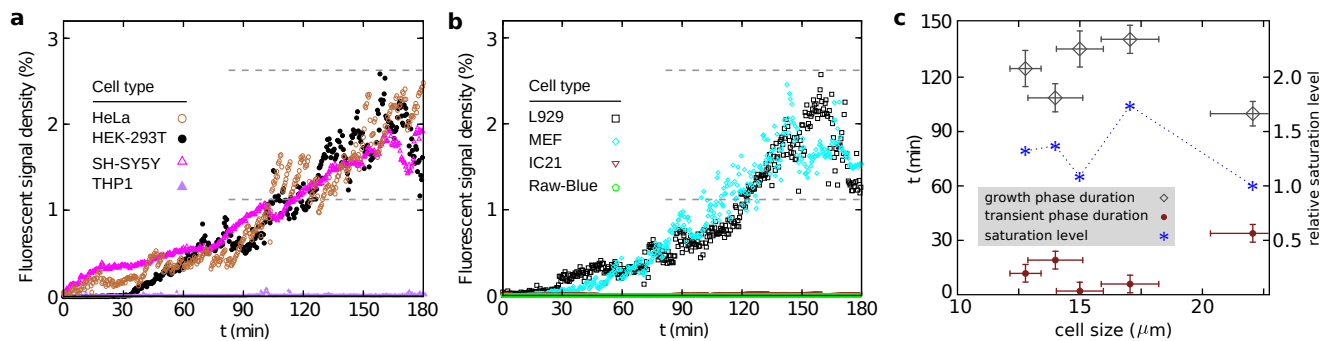}}
\caption{(a) Time evolution of the density of KDELR cargo signals at the surface 
of human cell lines after treatment with $\text{eGFP-RTA}^\text{E177D-HDEL}$. 
The optimal signal-to-noise ratio, scaled by the cell periphery size, 
is shown as a function of time. The dashed lines show the fluctuation 
range of the signal intensity at the steady state. (b) Similar to panel 
(a) but for mouse cell lines. (c) Duration of transient and growth 
regimes and the mean relative saturation level of signal density (compared 
to the largest analyzed cells) in the steady state versus the cell size.}
\label{Fig2}
\end{figure*}

Similar to the experiments with RTA, Hela cells develop KDEL receptor clusters 
in the presence of $\text{eGFP-RTA}^\text{E177D-HDEL}$ while no cluster 
formation is observed in control experiments using $\text{eGFP-RTA}^\text{E177D}$ 
lacking the ER-retention motif (Fig.\,\ref{Fig1}b). For other human cell lines, 
SH-SY5Y and HEK-293T, a qualitatively similar cluster development is observed, 
even though there are visible differences in the details of their temporal 
evolution. Surprisingly, differentiated human THP1 macrophages do not show 
any GFP clusters at the cell surface throughout the experiment. 

We also investigate cluster formation in the mouse cell lines. As can be 
seen in Fig.\,\ref{Fig1}c, the mouse fibroblast cell lines L929 and MEF exhibit 
a KDELR clustering similar to their human counterparts. Moreover, the macrophage 
cell lines IC21 and RAW-Blue likewise lack a KDELR clustering phenotype; both 
cell lines do not respond either to $\text{eGFP-RTA}^\text{E177D}$ or 
$\text{eGFP-RTA}^\text{E177D-HDEL}$ treatment. Therefore, we conclude 
that receptor clustering at the cell surface is not species specific 
and there are generally two types of cells: with or without ligand-induced 
KDELR cluster accumulation at the PM.

\begin{figure}
\centerline{\includegraphics[width=0.45\textwidth]{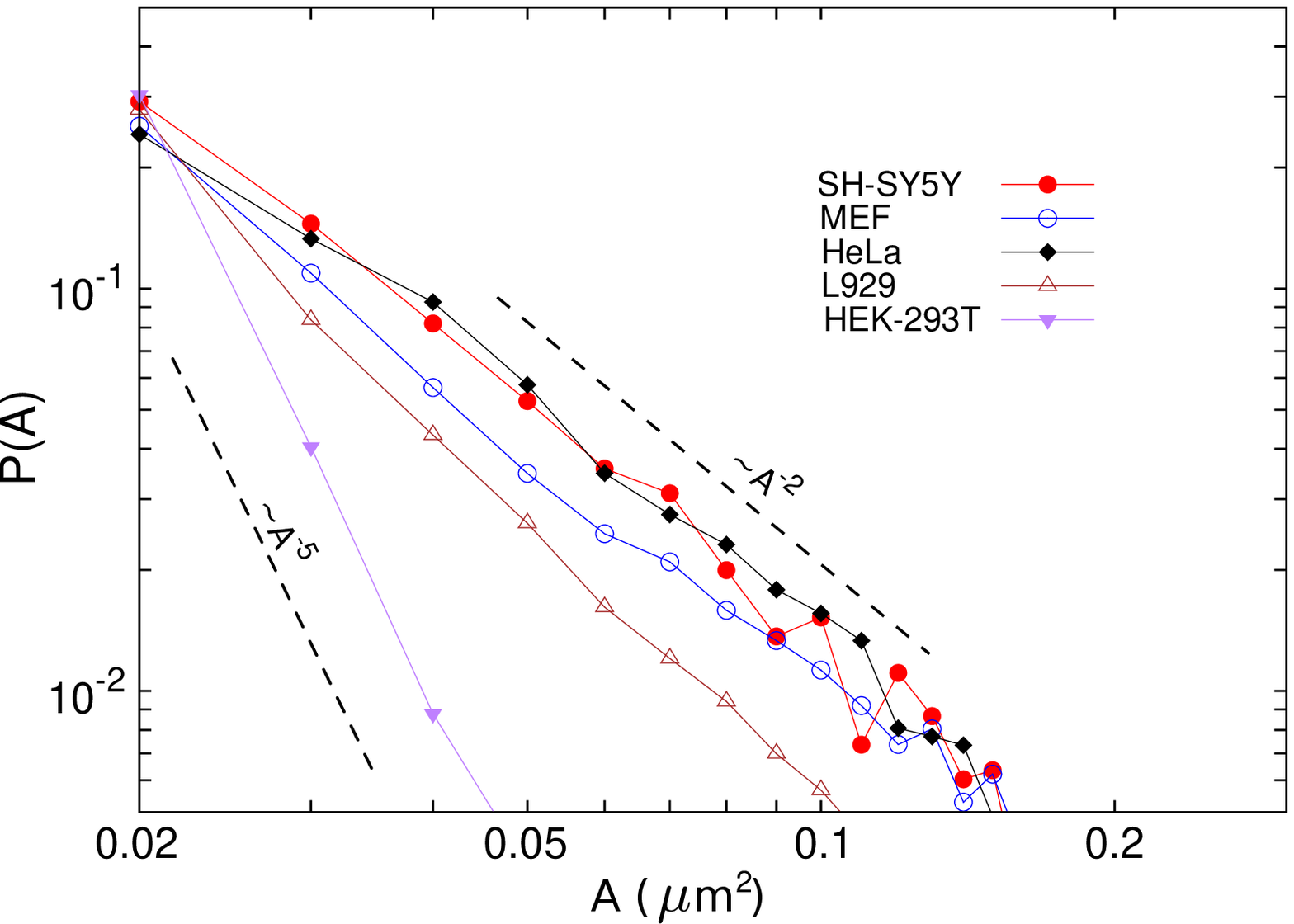}}
\caption{Cluster-size distribution of KDEL receptors at the surface of 
various cell types. The power-law exponent in the indicated cell lines 
varies from $\alpha{\simeq}2$ for HeLa and SH-SY5Y cells to $\alpha{\simeq}5$ 
for HEK cells.}
\label{Fig3}
\end{figure}

\begin{figure*}
\centerline{\includegraphics[width=0.47\textwidth]{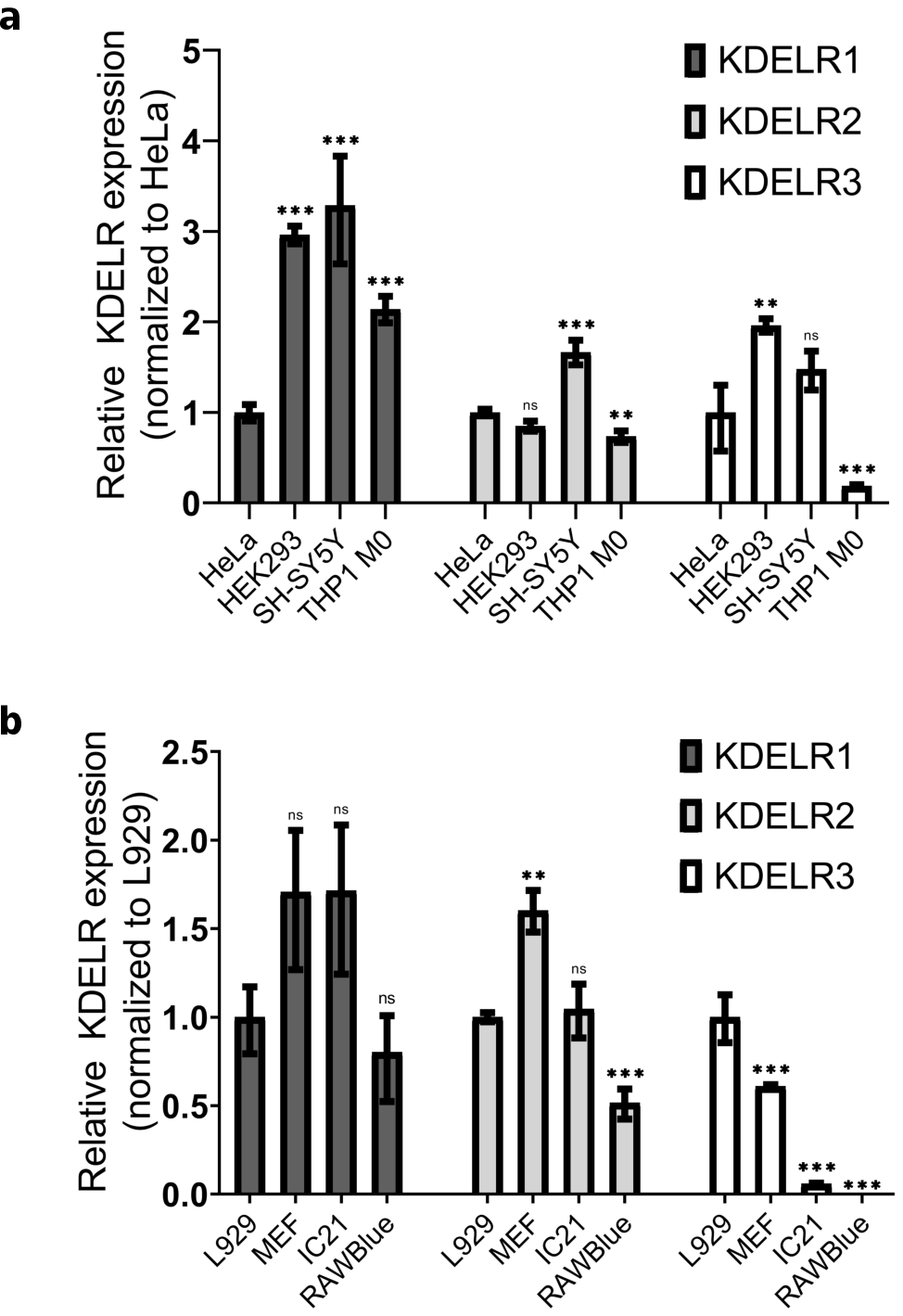}}
\caption{mRNA levels of three KDELR homologues in various (a) human 
and (b) mouse cell lines. Values in the indicated cell lines are scaled 
to the relative mRNA level of (a) HeLa and (b) L929 cells. Statistical 
significance is assessed by one-way ANOVA based on biological replicates 
and at sample sizes of n=3 (***p$\leq$0.001; **p$<$0.01; *p$<$0.05; ns, 
not significant).}
\label{Fig4}
\end{figure*}

Next, we take a closer look at the time evolution of the amount of fluorescent 
signals (i.e.\ clusters of $\text{eGFP-RTA}^\text{E177D-HDEL}$) at the cell 
surface. The cell lines showing cluster formation follow a similar temporal 
evolution as shown in Fig.\,\ref{Fig2}: the system initially remains almost 
inactive for a relatively short time (transient regime). Then, the clustering 
process speeds up with an increasing slope (exponential growth regime), until 
it eventually reaches a non-equilibrium steady state where the signal density 
fluctuates around a mean value due to the interplay between the stochastic 
loss of receptors via endocytosis and gain by exocytosis events (steady-state 
regime). Although the overall clustering process is similar in these cell 
types, the details of the temporal evolution of signal density differ from 
one cell to another. In order to clarify if the observed diversity is originated 
from the cell-size differences, we categorize the analyzed cells based on 
their sizes. To this aim, we approximate the mean cell size by the 
diameter of a circle with the same area as the cell. Figure\,\ref{Fig2}c 
shows that there is no systematic dependence between the characteristics of 
the three regimes and the cell size. Presumably, the endocytosis and exocytosis 
rates are the main influential factors in the formation and growth of receptor 
clusters. 

\subsection*{Cluster size distribution} 
It was previously shown that the size distribution of the receptor clusters at 
the surface of HeLa cells decays as a power-law, indicating that there are 
preferred arrival sites at the plasma membrane  \cite{Becker16b}. Our analysis 
of the size of clusters in different cell types reveals that the cluster-size 
distribution in all cases nearly follows an algebraic form $P(A){\sim}A^{-\alpha}$ 
(Fig.\,\ref{Fig3}). The decay exponent is around $\alpha{=}2$ for HeLa and SH-SY5Y 
and $\alpha{>}2$ for MEF, L929, and HEK cells. The exponents greater than 2 may 
evidence for a less efficient clustering process which prevents the formation 
of large clusters. 

\subsection*{KDELR mRNA levels are not correlated with cluster formation.} 
As a step towards identifying the responsible factors for the observed 
non-clustering phenotype of macrophage cell lines, we perform RT-qPCR 
experiments and compare the intracellular mRNA levels of three KDELR 
homologues to see if there are differences on the transcriptional level 
between macrophage-derived and other cells including fibroblast mouse cell 
lines (see Fig.\,\ref{Fig4}). The consistency of our data with the previous 
RT-qPCR results of HeLa \cite{Raykhel07} and HEK-293T and SH-SY5Y \cite{Trychta18} 
cells verifies the reliability of our measurements. We find no correlation 
between the mRNA level of KDELR1 or KDELR2 and the observed clustering 
differences. mRNA levels of KDELR3 in all macrophage cell lines are 
significantly lower than in cluster-forming cells. However, the basal mRNA 
expression of KDELR3 is also extremely low compared to KDELR1/2 in all cell 
lines as shown in {\it Suppl.\,Fig.\,S1}. Therefore, it is unlikely that the 
low mRNA level of KDELR3 is responsible for the missing receptor clustering 
phenotype seen in macrophages. 

\section*{Numerical approach}
The results presented in the previous section raise the natural 
question: what is the difference between macrophage and other cell types? 
The formation of receptor clusters at the cell surface is a non-equilibrium 
process which ultimately reaches a steady state when the loss of receptors 
due to endocytosis is balanced with the gain by recycling them. We expect 
that an extremely high endocytosis or low exocytosis rate can practically 
prevent the formation of clusters. It is, however, unclear how large/small 
such rates have to be compared to those of cluster-forming cells. Since 
the adequate staining or detection of KDELRs via Western analysis or 
immunofluorescence is impossible due to a lack of suitable isoform-specific 
antibodies, we do not have the opportunity to biochemically quantify the 
amount of PM-localized KDELRs in different 
cell lines. 

\subsection*{The model} 
\begin{figure}
\centerline{\includegraphics[width=0.45\textwidth]{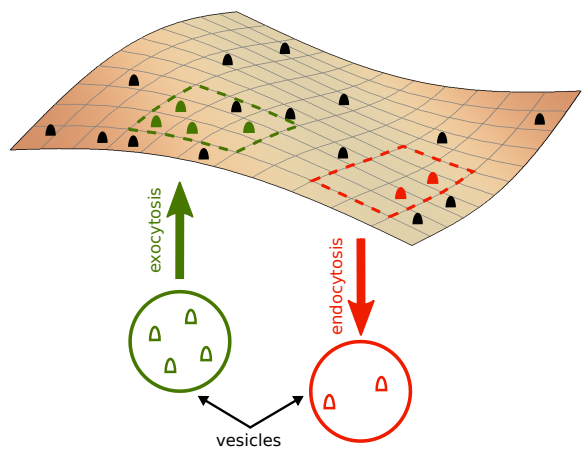}}
\caption{Schematic illustration of the receptor cycling model. Black, 
green, and red full symbols represent, respectively, the receptors 
which survive or will be added or eliminated in the next time step. 
Dashed lines indicate the affected zones by endo/exocytosis events.}
\label{Fig5}
\end{figure}

To gain an insight into the interplay between endocytosis and exocytosis 
rates, we consider a stochastic process of internalization and vesicle 
arrival events to model the loss and gain of receptors at the cell surface. 
By ignoring the self-amplification effects for simplicity, we model the 
process in the following way: {\it (i) cell surface}: The membrane is 
modeled as a square lattice with periodic boundary conditions to mimic 
the closed cell surface. The lattice grid size is chosen to be 5 nm, 
which is in the order of magnitude of the KDEL receptor size 
\cite{Brauer19,Reddy10,Guo99}. Each lattice site is occupied by a 
liganded/unliganded receptor or remains empty (Fig.\,\ref{Fig5}). 
We consider a system of size $L{=}10\,\mu\text{m}$, which is 
the approximate size of our smallest analyzed cells. {\it (ii) 
endocytosis events}: We assume that the positions of endocytosis 
events are uncorrelated in space and time. Thus, a random site is 
chosen as the center of the endocytosis event at each time step. 
The extent of the region affected by the endocytosis event is chosen 
randomly within 5 to 10 lattice sites around the center site, since 
the typical size of clathrin-coated vesicles is $\sim$50-100 nm 
\cite{Agrawal10}. All the existing receptors within the affected 
region are eliminated (i.e.\ internalized). {\it (iii) exocytosis 
events}: The center of the target zone is randomly chosen from a 
multiple-peaked Gaussian distribution, with the peaks representing 
the places where microtubules approach the plasma membrane. The size 
of the arrival vesicle is chosen similarly to the endocytosis events. 
The random number of receptors carried by the arrival vesicle ranges 
from zero to the maximum capacity of the vesicle. These receptors 
are randomly distributed in the affected region upon availability 
of empty sites. 

\begin{figure*}
\centerline{\includegraphics[width=0.99\textwidth]{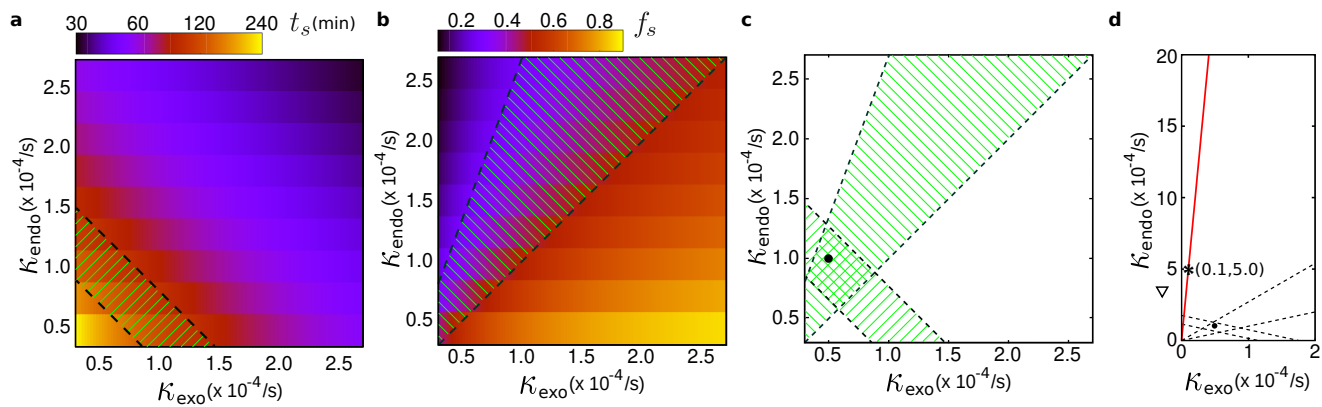}}
\caption{(a) Characteristic time $t_s$ and (b) relative saturation 
level $f\!_s$ in the ($\kappa_{\text{endo}}$, $\kappa_{\text{exo}}$) 
phase space. Single-hashed regions in panels (a) and (b) correspond 
to the values of $\kappa_{\text{endo}}$ and $\kappa_{\text{exo}}$ 
which result in $t_s$ or $f\!_s$ values experimentally observed for 
cluster-forming cells. (c) Double-hashed region shows the approximate 
extent of $\kappa_{\text{endo}}$ and $\kappa_{\text{exo}}$ rates in 
cell lines showing receptor clustering. The full circle represents 
the reference point (see text). (d) Similar to panel (c) but with 
different plot range. The solid line corresponds to the threshold 
line $\kappa_{\text{endo}}{=}(-1{+}1{/}f\!_s^{\,c})\kappa_{\text{exo}}$, 
representing the unset of undetectable saturation level in experiments.}
\label{Fig6}
\end{figure*}

\subsection*{Formation and evolution of clusters} 
Starting from an initially empty membrane, surface density evolves 
and eventually reaches a non-equilibrium steady-state level with 
relatively large fluctuations. We characterize the steady state 
with the saturation time $t_s$ (i.e.\ the characteristic time 
needed to reach $1{-}\frac1e$ fraction of the mean saturation 
density) and the saturation level $f\!_s$ (defined as the fraction 
of the total receptors of the cell that are located on the cell 
surface, thus, $f\!_s{\in}[0,1]$). Figure\,\ref{Fig6} shows how 
receptor clustering depends on the endocytosis $\kappa_{\text{endo}}$ 
and exocytosis $\kappa_{\text{exo}}$ rates (i.e.\ the effective 
receptor elimination and arrival rates in our Monte Carlo 
simulations). The characteristic time $t_s$ varies by several 
orders of magnitude depending on the choice of $\kappa_{
\text{endo}}$ and $\kappa_{\text{exo}}$. To determine the subset 
of the ($\kappa_{\text{endo}}$, $\kappa_{\text{exo}}$) phase 
space which is covered by cluster-forming cells, one can 
solve the master equation $\dot{f}{=}-\kappa_{\text{endo}}f{+}
\kappa_{\text{exo}}(1{-}f)$ for the saturation level $f$ 
to obtain its temporal evolution $f(t)=f\!_s+(f\!_{_0}{-}f\!_s)
\,\text{e}^{-t{/}t_s}$,with $f\!_{_0}$ being the initial saturation 
level (taken to be zero in our simulations) and $t_s{=}1{/}
(\kappa_{\text{endo}}{+}\kappa_{\text{exo}})$ being the characteristic 
time. Denoting the minimum and maximum duration of the exponential 
growth regime in experiments, respectively, with $t_s^{\text{min}}$ 
and $t_s^{\text{max}}$, then the lines $\kappa_{\text{endo}}
{+}\kappa_{\text{exo}}{=}1{/}t_s^{\text{min}\,\text{(max)}}$ 
define the dashed lines in Fig.\,\ref{Fig6}a as the lower and 
upper borders of the endo/exocytosis rates. Furthermore, from 
the above master equation it can be seen that the relative 
saturation level is also given by the endo/exocytosis rates 
as $f\!_s{=}\kappa_{\text{exo}}{/}(\kappa_{\text{endo}}{+}
\kappa_{\text{exo}})$. Figure~\ref{Fig6}b shows that the 
saturation level reduces with increasing the endocytosis 
and decreasing the exocytosis rates (upper-left corner). 
Assuming that $f\!_s$ fluctuates in the experimental data 
within $[f\!_s^{\;\text{min}}, f\!_s^{\;\text{max}}]$, we can 
determine the borders of the experimentally accessible region 
of the ($\kappa_{\text{endo}}$, $\kappa_{\text{exo}}$) phase 
space via the lines $\kappa_{\text{endo}}{=}(-1{+}
\frac{1}{f\!_s^{\;\text{min}\,\text{(max)}}})
\kappa_{\text{exo}}$. While we know from our experiments 
that $f\!_s^{\;\text{max}}{/}f\!_s^{\;\text{min}}{\sim}1.85$, 
the absolute value of $f\!_s^{\;\text{min}}$ or $f\!_s^{\;
\text{max}}$ is unknown as we have no possibility to 
estimate the fraction of the total receptors of the cell 
that are located on the cell surface. Let us, for example, 
suppose that at most half of the receptors present at the 
plasma membrane in the steady state (i.e.\ $f\!_s^{\;\text{max}}
{=}0.5$ and thus $f\!_s^{\;\text{min}}{\simeq}0.27$). This 
results in the two dashed lines in Fig.\,\ref{Fig6}b. Note 
that changing the values of $f\!_s^{\;\text{min}}$ or $f\!_s^{\;
\text{max}}$ will change the slopes of the dashed lines in 
Fig.\,\ref{Fig6}b but does not qualitatively affect our 
following argument and conclusions. By combining the hashed 
regions in panels (a) and (b) of Fig.\,\ref{Fig6}, the double-hashed 
zone in Fig.\,\ref{Fig6}c is obtained which displays the approximate 
range of $\kappa_{\text{endo}}$ and $\kappa_{\text{exo}}$ rates 
for the cell lines showing cluster formation (i.e.\ HeLa, SH-SY5Y, 
HEK-293T, L929, and MEF cells). We choose a representative reference 
point in the bulk of this zone with $\kappa_{\text{endo}}{=}10^{-4}$/s 
and $\kappa_{\text{exo}}{=}0.5{\times}10^{-4}$/s. Assuming that the 
relative saturation level $f\!_s$ should be greater than a threshold 
(e.g.\ $f\!_s^{\,c}{=}0.02$) to be detectable in our experimental measurements, 
we calculate how far one should move from the reference point along 
the $\kappa_{\text{endo}}$ or $\kappa_{\text{exo}}$ axis to reach 
the undetectable threshold value $f\!_s^{\,c}$. We find that either 
an extremely high endocytosis rate $\kappa_{\text{endo}}{\sim}24.5
{\times}10^{-4}$/s (approximately 25 times higher than the reference 
$\kappa_{\text{endo}}$ value) or an extremely low exocytosis rate 
$\kappa_{\text{exo}}{\sim}0.02{\times}10^{-4}$/s (approximately 25 
times lower than the reference $\kappa_{\text{exo}}$ value) is 
required to have an undetectable receptor clustering saturation 
level. Using the former (latter) extreme rate and keeping the other 
rate fixed at its reference value, the system will quickly converge 
to the steady state in nearly seven minutes (a few seconds). While 
such extreme differences between the endo/exocytosis rates of macrophages 
and cluster-forming cells might be unexpected, it should be noted 
that any other combination of the rates satisfying $\kappa_{\text{endo}}
{=}(-1{+}1{/}f\!_s^{\,c})\kappa_{\text{exo}}$ (solid red line in 
Fig.\,\ref{Fig6}d) would also be a solution. For instance, a 
system with $\kappa_{\text{endo}}{=}5{\times}10^{-4}$/s and 
$\kappa_{\text{exo}}{=}0.1{\times}10^{-4}$/s (see the star shown 
in Fig.\,\ref{Fig6}d with a 5-fold difference with the coordinates 
of the reference point along each direction) converges to 
$f\!_s^{\,c}$ in nearly half an hour. Therefore, having only 
a few fold faster endocytosis rate together with a few fold 
slower exocytosis rate (compared to the reference point) can 
lead to an undetectable level of surface receptors at the 
steady state. Note that in the above discussion we have 
assumed that the surface density of receptors $n$ is nearly 
the same for endo- and exocytosis. This assumption is critical 
at low densities because at least one receptor has to be present 
in order to trigger endocytosis. Since the probability of 
endocytosis depends on the size of local receptor clusters, the 
density of receptors in endo- and exocytosis might be different. 
We can account for this effect by introducing an effective 
endocytosis rate $\kappa'_{\text{endo}}{=}n\,\kappa_{\text{endo}}$ 
containing the combined effects of internalization rate and surface 
occupation density. Then the above discussion remains valid for 
the effective endocytosis rate $\kappa'_{\text{endo}}$. 

\subsection*{Power-law decay of the cluster-size distribution} 
Towards understanding the origin of the diversity observed in the 
power-law exponent $\alpha$ of the cluster-size distribution in 
experiments (Fig.\,\ref{Fig3}), we consider a simple receptor 
aggregation process in which the preferential attachment of the 
newly arrived receptor to the existing clusters occurs with a 
given probability $\beta$. Indeed, $\beta$ represents the efficiency 
of the directed transport to preferred docking sites on the membrane. 
Thus, the probability to attach to a cluster of size $A_i$ is 
$\displaystyle P(A_i){=}\frac{\partial A_i}{\partial t}{=}
\beta\frac{A_i}{\sum_j A_j}$, where the sum runs over all existing 
clusters. Since $\sum\limits_j A_j$ grows linearly with time, 
we have $\displaystyle \frac{\partial A_i}{\partial t}{=}
\beta\frac{A_i}{t}$ leading to $A_i(t){=}(\frac{t}{t_i})^{\beta}$, 
with $t_i$ being the initiation time of cluster $i$. Then the 
cumulative probability that a cluster is smaller than $A$ can be 
obtained as $F[A_i(t){<}A]=F(t_i{>}\frac{t}{A^{1{/}\beta}})
=1{-}\frac{1}{A^{1{/}\beta}}$, where we assumed that $F(t_i)$ 
has a constant probability density, i.e.\ $F(t_i){=}\frac{1}{t}$. 
Finally, the cluster-size distribution can be derived as $P(A)
=\frac{\partial F[A_i(t){<}A]}{\partial A}=A^{-(1{+}1{/}\beta)}$. 
This suggests that the power-law exponent $\alpha$ is related to 
the efficiency of the preferential attachment via $\alpha{=}
1{+}1{/}\beta$. For $\beta{=}1$, every arriving receptor chooses 
the preferred hot spots on the plasma membrane, leading to 
$\alpha{=}2$. However, $\alpha$ grows as the probability $\beta$ 
to attach to the preferred docking sites on the membrane decreases. 
For instance, one obtains $\alpha{=}5$ for $\beta{=}1{/}4$. 
An inefficient preferential attachment process slows down the 
growth rate of clusters and it is less likely that large 
clusters form on short time scales; as a result, the tail 
of the cluster-size distribution decays faster. 

\section*{Disscussion and conclusions}
\noindent In summary, we studied KDEL receptor clustering 
at the plasma membrane of different mouse and human cell 
lines via live cell imaging. We verified that the cluster 
formation is independent of species identity but there 
are cell-type differences: while no cluster formation 
is observed for macrophage cell lines from both species, 
other cell types such as fibroblast mouse cells develop 
clusters in a qualitatively similar manner. 

Assuming that the KDELR internalization and recycling process 
exists in all cell lines including the macrophages, our numerical 
analysis suggests a few possible scenarios to prevent the formation 
of receptor clusters at macrophage cell surfaces: (i) the total 
number of receptors is too low in macrophages such that the arrival 
vesicles distribute too few receptors on the cell surface; (ii) 
one of the rates (either endocytosis or exocytosis) in macrophages 
differs considerably from that of cluster-forming cells, while 
the other rate behaves similarly to other cell types. According 
to this scenario, the endocytosis (exocytosis) events in macrophages 
should occur with at least a 25-fold higher (lower) rate compared 
to cluster-forming cells; (iii) a combination of faster 
endocytosis and slower exocytosis rates is responsible for the 
missing receptor clustering phenotype seen in macrophages. In 
this case, moderate differences compared to cluster-forming cells, 
such as a 5-fold higher endocytosis rate together with a 5-fold 
lower exocytosis rate, suffice to prevent the formation of receptor 
clusters at macrophage cell surfaces; and (iv) the surface density 
of KDELRs in macrophages may differ for endo- and exocytosis, compared 
to cluster-forming cells. 

The first scenario is already disproved indirectly by our RT-qPCR 
results, verifying that the levels of KDELR1/2 mRNA in cluster-forming 
cell lines are similar to the cells forming no clusters. Although 
mRNA level of KDELR3 is lower in macrophages, we expect that 
this receptor type plays a rather insignificant role in general 
due to the extremely low expression level of KDELR3 compared to 
KDELR1/2 (see {\it Suppl.\,Fig.\,S1}).

As a part of the immune system, macrophages serve as professional 
phagocytotic cells and are specialized to detect and quickly 
eliminate pathogen particles such as cell debris or bacteria 
\cite{Elhelu83,Hirayama17}. The internalization rate of macrophages 
strongly depends on the type of particle, ranging from a half-life 
of a couple of seconds for fast phagocytotic events \cite{Magnusson89} 
to minutes or even hours for ligand/receptor endocytosis \cite{Mellman83,
Ward90}. There are however other cell types which perform clathrin-driven 
receptor endocytosis with a similar rate and internalize surface 
bound ligands in a few minutes \cite{Sorkin96,SchneiderBrachert04,
Ciechanover83}.

An extremely high endocytosis rate implicates a high ligand uptake 
rate. Subsequently, intracellular KDELR/ligand signals should be 
visible over the long imaging period of 3 h. We however observe 
no GFP signals in macrophage cell lines. Rapid lysosomal degradation 
and the associated deprivation of the GFP fluorescence in the model 
cargo is also unlikely, because the interaction with KDELRs should 
mainly prevent the ligand transport in this organelle and foster 
its targeted retrograde transport into the ER. 

A very low exocytosis rate of KDELRs to the PM could be also 
a possible explanation for the observed non-clustering phenotype. 
Nevertheless, the quantification of PM-localized KDELRs is so far 
impossible due to the lack of suitable antibodies for immunofluorescence 
studies. Moreover, overexpression of tagged KDELRs is in principle 
possible but the overload of the natural ER retention system may 
dramatically affect the results leading to misinterpretations.

A complete absence of PM-localized KDELRs could be another explanation 
for the observed phenotype. It is possible that macrophage cell lines 
have an active mechanism to prevent PM-transport of KDELRs or lacking 
specific cellular components (i.e.\ proteins and/or signaling pathways) 
required for proper cell surface transport or ligand binding recognition. 
In such a case, macrophages cannot respond to the external applied ligand 
and receptor clusters do not form. Our results thus call for systematic 
studies to better understand the internal mechanisms of receptor clustering 
and to clarify the differences between macrophages and other cell types, 
e.g.\ in their endo/exocytosis rates or in the total amount of PM-localized 
KDELRs. To this aim, better antibodies and tools are required for quantitative 
comparisons. Understanding why macrophages do not form ligand/receptor 
clusters could also shed light on how these cells achieve an efficient 
immune response and interact with tissue cells containing KDELRs at 
their cell surface. \\

\subsection*{Acknowledgements}
We thank Prof.\ Marc Schneider for the possibility of performing the live 
cell imaging experiments in his lab. We further thank Profs.\ Ar\'anzazu 
del Campo and Gernot Geginat for providing MEF, L929 and IC21 cell lines. 
This work was funded by the Deutsche Forschungsgemeinschaft (DFG) through 
Collaborative Research Center SFB 1027 (Projects A6, A7, A8).

\subsection*{Author contributions statement}
B.B., M.J.S., M.R.S.\ and L.S.\ designed the research. A.B.\ and B.B.\ 
performed the experiments. A.B.\ and M.R.S.\ analyzed the experimental 
data. M.R.S.\ performed the simulations and numerical analysis. All 
authors contributed to the interpretation of the results. B.B.\ and 
M.R.S.\ wrote the manuscript. B.B. and M.R.S. contributed equally to 
this work. Correspondence should be addressed to B.B (bjoern\_becker2@gmx.de) 
or M.R.S. (shaebani@lusi.uni-sb.de).

\subsection*{Competing financial interests} 
The authors declare no competing financial interests.

\subsection*{Supplementary information} 
Supplementary information accompanies this paper, including one figure 
and two tables.

\newpage

\begin{center}
\includegraphics[width=0.6\textwidth]{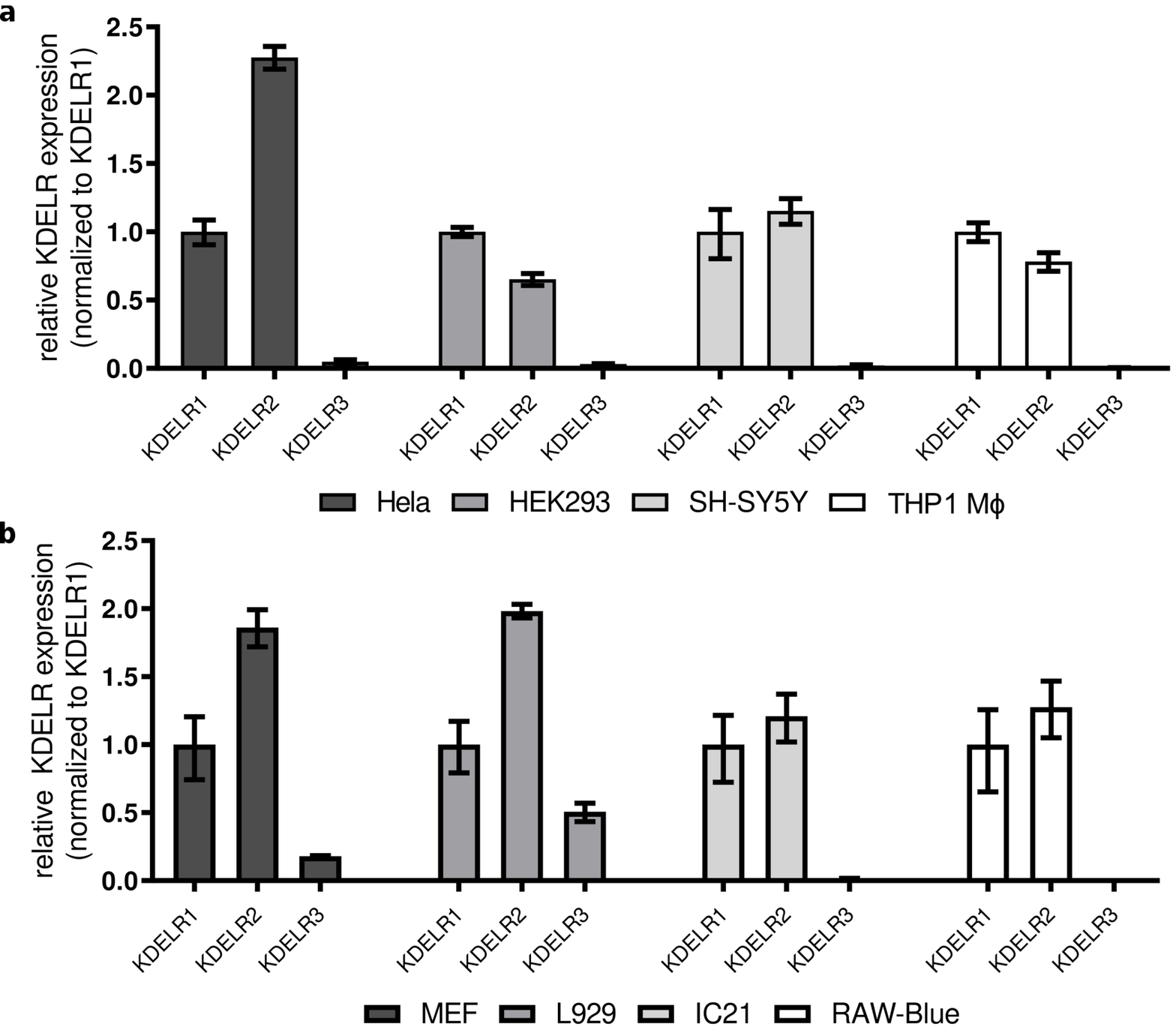}
\end{center}
{\bf Supplementary Figure S1: }{\normalsize Relative KDELR1-3 mRNA 
levels determined with RT-qPCR (n=3) in (a) human cells (HeLa, 
Hek-293T, SH-SY5Y, and THP1) and (b) mouse cells (L929, MEF, IC21, 
and RAW-Blue). mRNA levels were normalized to the human or mouse 
house keeping gene GAPDH and KDELR1 mRNA level was set to $100\%$.}

\begin{table*}
\begin{tabular*}{\hsize}{l}
{\bf Supplementary Table S1.} Primer sequences used in this study \cr
\end{tabular*}
\begin{tabular*}{\hsize}{@{\extracolsep{\fill}}lll}
\hline \noalign{\smallskip}
Primer name&5'-3' sequence \cr \noalign{\smallskip}
\hline \noalign{\smallskip}
5'-$\text{RTA}^\text{E177D}$&GAATTCGGATCCATGATATTCCCCAAACAATACCCAATTATAAACTTTACC \cr \noalign{\smallskip}
3'-$\text{RTA}^\text{E177D}$&AAGCTTGTCGACTTAATGATGATGATGATGATGAAACTGTGACGATGGTGGAGGTGC \cr \noalign{\smallskip}
3'-$\text{RTA}^\text{E177D-HDEL}$&AAGCTTGTCGACTTACAGTTCATCATGATGATGATGATGATGATGAAACTGTGACGATGGTGGA \cr \noalign{\smallskip}
&GGTGC \cr \noalign{\smallskip}
\hline \bigskip
\end{tabular*}
\end{table*}

\begin{table*}
\begin{tabular*}{\hsize}{l}
{\bf Supplementary Table S2.} Nucleotide sequences for qPCR primers (ordered 
from Invitrogen) \cr
\end{tabular*}
\begin{tabular*}{\hsize}{@{\extracolsep{\fill}}lll}
\hline \noalign{\smallskip}
Target&Forward&Reverse \cr \noalign{\smallskip}
\hline \noalign{\smallskip}
Human KDELR1 NC\,000019&CACAGCCATTCTGGCGTTCCTG&CCATGAACAGCTGCGGCAAGAT \cr \noalign{\smallskip}
Human KDELR2 NC\,000007&CTGGTCTTCACAACTCGTTACCTGGATC&CAGGTAGATCAGGTACACTGTGGCATAGG \cr \noalign{\smallskip}
Human KDELR3 NC\,000022&CTTCTGGTCCCAGTCATTGGCCT&GGGGCAGGATAGCCACTGATTCC \cr \noalign{\smallskip}
Human GAPDH NC\,000012&TTCGACAGTCAGCCGCATCT&GCCCAATACGACCAAATCCGTT \cr \noalign{\smallskip}
Mouse KDELR1 NC\,000073&GTGGTGTTCACTGCCCGATA&AACTCCACCCGGAAAGTGTC \cr \noalign{\smallskip}
Mouse KDELR2 NC\,000071&TGGTCTTCACGACTCGCTAC&AGGTACACCGTGGCATAGGA \cr \noalign{\smallskip}
Mouse KDELR3 NC\,000081&CTTCATCTCCATCTACAACACAGTG&CTCCAGCCGGAATGTGTCAT \cr \noalign{\smallskip}
Mouse GAPDH NC\,000072&GAGAGTGTTTCCTCGTCCCG&TCCCGTTGATGACAAGCTTCC \cr \noalign{\smallskip}
\hline \bigskip
\end{tabular*}
\end{table*}

\end{document}